\documentclass[%
 reprint,
 amsmath,amssymb,
 aps,
]{revtex4-2}

\usepackage{graphicx}
\usepackage{dcolumn}
\usepackage{bm}

\usepackage{graphicx}
\usepackage{booktabs}
\usepackage{svg}
\usepackage{subcaption}

\usepackage{ragged2e} 

\usepackage{adjustbox}   
\renewcommand{\arraystretch}{1}  
\usepackage{tabularx}
\svgpath{{figures/}} 

\begin{document}

\preprint{APS/123-QED}

 \title{Frequency-matching quantum key distribution}

\author{Hao-Tao Zhu$^{1,2,3,4}$,
	Yizhi Huang$^{5}$,
	Abdullah Rasmita$^{1,2,3,4}$,
	Chao Ding$^{1,2,3,4}$,
	Xiangbin Cai$^{1,2,3,4}$,
	Haoran Zhang$^{1,2,3,4}$,
	Xiongfeng Ma$^{5,*}$,
	Weibo Gao$^{1,2,3,4,}$}
 \altaffiliation[]{xma@tsinghua.edu.cn; wbgao@ntu.edu.sg}

\affiliation{%
$^{1}$School of Electrical and Electronic Engineering,
Nanyang Technological University, Singapore 639798, Singapore\\
$^{2}$Division of Physics and Applied Physics, School of Physical and Mathematical Sciences,
Nanyang Technological University, Singapore 637371, Singapore\\
$^{3}$Centre for Quantum Technologies, National University of Singapore, Singapore\\
$^{4}$Quantum Science and Engineering Centre (QSec),
Nanyang Technological University, Singapore\\
$^{5}$Center for Quantum Information, Institute for Interdisciplinary Information Sciences,
Tsinghua University, Beijing 100084, China
}%


%

\date{\today}

\begin{abstract}
Quantum key distribution (QKD) enables information-theoretically secure communication against eavesdropping. However, phase instability remains a challenge across many QKD applications, particularly in schemes such as twin-field QKD and measurement-device-independent QKD. The most dominant source of phase fluctuation arises from the frequency offset between independent lasers. To overcome this limitation, we introduce a classical photodiode-based active compensation scheme for laser frequency mismatches and incorporate it into mode-pairing QKD protocol. We have achieved an error rate approaching the theoretical limit and surpassing the linear key-rate bound over a fiber distance of 296.8 km. This approach provides a practical solution for frequency matching between independent lasers and can be extended to other fields requiring precise phase stabilization.

\end{abstract}

\maketitle


\section{\label{sec:level1}Introduction}

Quantum key distribution (QKD), a fundamental pillar of quantum technology, provides an information-theoretically secure means of communication, immune to the sophisticated eavesdropping techniques \cite{bennett1984quantum,ekert1991Quantum,lo1999unconditional,shor2000simple,10.5555/2011586.2011587,scarani2009security,brassard2000limitations,bennett1992quantum,renner2005information,ribezzo2023deploying,renner2008security,devetak2005distillation}. Since its first proof-of-concept demonstrations \cite{Bennett1992experimental}, QKD has rapidly progressed from theory to practice. Significant achievements, such as satellite-based QKD enabling secure intercontinental communication \cite{Liao2018Satellite,bedington2017progress} and terrestrial quantum communication \cite{Chen2021integrated,martin2024madqci,liu2023experimental}, underscore its transformative impact.

In various QKD protocol schemes, accurate estimation or compensation of the phase is essential for optimizing protocol performance\cite{lucamarini2018overcoming,zeng2022mode}. Two primary approaches have been explored. The first approach inserts reference pulses among the QKD signal pulses, where the phase difference is estimated from interference measurements using single-photon detectors, analyzed via maximum likelihood estimation\cite{zhu2023experimental} or Fast Fourier Transform (FFT) method\cite{li2023twin,zhang2025experimental,shao2025high}. The phase information of the QKD pulses is then inferred by fitting or extrapolating from the reference light detection. The second approach directly stabilizes the phase of the pulses during transmission by employing ultra-stable laser technologies, such as optical cavities\cite{zhou2023experimental} or acetylene-stabilized lasers\cite{chen2024twin,ge2024post}, which enables real-time phase stabilization without the need for additional reference measurements.

The first approach requires superconducting single-photon detectors with sufficiently high-count rates. Also, it reduces the duty cycle of QKD, demands additional modulation of reference light, and introduces significant computational overhead, making practical deployment challenging. Furthermore, the probabilistic detections and dead time of single-photon detectors necessitate more detection events and longer acquisition times to achieve accurate phase estimation. By contrast, the second approach provides stable and precise phase estimation, but its reliance on costly and technically demanding infrastructure imposes significant barriers to large-scale deployment in quantum networks.

To address these issues, we propose a high-precision phase-estimation method that correlates interference measurements from a classical photodiode. The photodiode records the beat note of two frequency-matching classical beams. This signal directly reveals their instantaneous frequency offset and the phase evolution relevant to the quantum signal. Due to its gigahertz-level bandwidth, the photodiode enables accurate real-time tracking of frequency offsets while tolerating slow drifts in the optical paths.

\begin{figure}[htbp]
	\includegraphics[width=8cm]{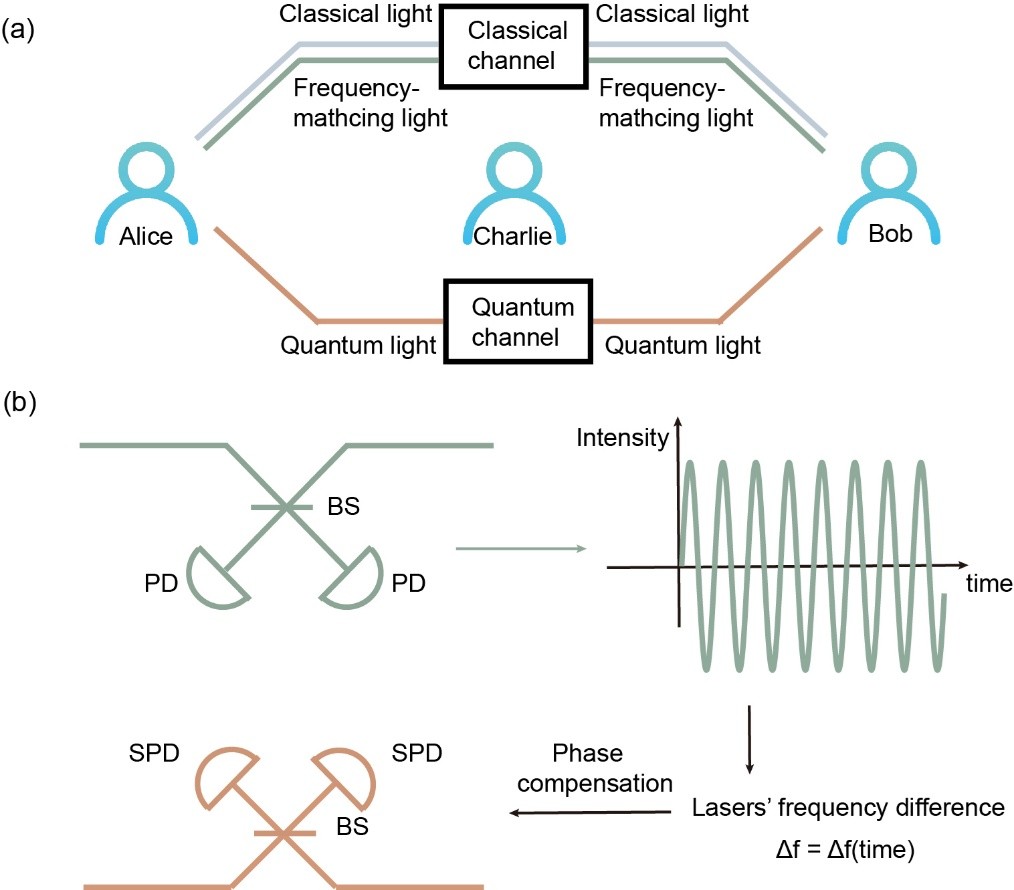}
    \captionsetup{justification=justified,singlelinecheck=false}
	\caption{\justifying Schematic of the frequency-matching scheme. (a) Alice and Bob transmit quantum light through a quantum channel to an intermediate node, Charlie, while classical channels carry synchronization and frequency-matching light. (b) Photodiode beat-note measurement for laser frequency-offset extraction, correlated with single-photon detection events for precise phase-difference estimation. Abbreviations: PD, photodiode; SPD, single-photon detector; BS, beam splitter.
	} \label{fig:Mainidea}
\end{figure}

As shown in Fig. \ref{fig:Mainidea}, the photodiode trace is synchronized with single-photon detection events. Since the classical frequency-matching beams co-propagate with the quantum pulses, the measured offset directly reflects the channel phase evolution. With the strong light intensity, this enables quantum phase estimation without single-photon detectors as used in Refs.~\cite{zhu2023experimental,li2023twin,zhang2025experimental,shao2025high} and obtain a more accurate phase stabilization.

\begin{figure*}[htbp]
	\includegraphics[width=18cm]{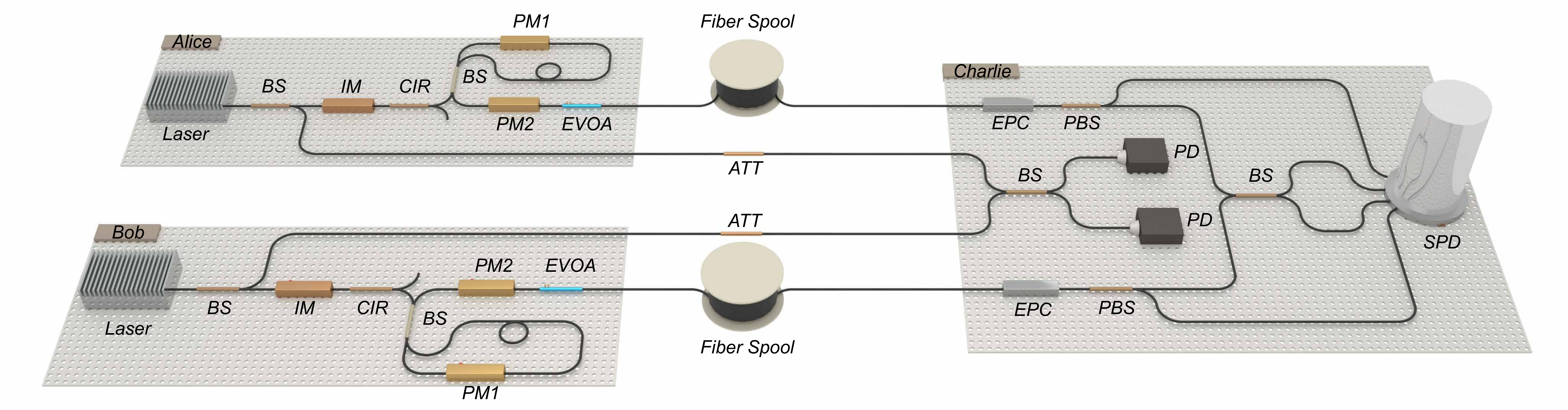}
	\caption{\justifying Experimental Setup. Alice and Bob employ narrow-linewidth lasers as light sources, splitting the output into two beams. One beam is directed to Charlie for interference, which facilitate frequency matching. The second beam undergoes quantum light modulation, including chopping and decoy state modulation via an intensity modulator and a Sagnac loop, comprising a circulator, beam splitter, phase modulator, and optical fibers. A phase modulator applies 16 distinct phase levels for phase randomization. The modulated quantum light pulses are transmitted to Charlie, where polarization feedback aligns the polarization of the two beams. Interference is then performed using a beam splitter. The resulting signals from both PDs and single-photon detectors are processed simultaneously for post-processing, ultimately generating secure keys. BS: beam splitter; IM: intensity modulator; CIR: circulator; PM: phase modulator; EVOA: electrical variable optical attenuator; EPC: electric polarization controller; PBS: polarizing beam splitter; PD: photodiode; SPD: single-photon detector.}
	\label{fig:expsetup}
\end{figure*}



\section{Mode-pairing QKD protocol}
Firstly, we outline the key steps of the scheme. Alice and Bob independently
prepare coherent states. Alice generates a state
\(
\left|\sqrt{\mu^a}\,e^{i\phi^a}\right\rangle
\),
where the intensity \( \mu^a \) is randomly chosen from
\(\{0, \nu^a, \mu^a\}\) and the phase \( \phi^a \) is selected from
\(\{0, 2\pi/D, \ldots, 2\pi(D-1)/D \}\).

Bob follows the same process, preparing
\(
\left|\sqrt{\mu^b}\,e^{i\phi^b}\right\rangle
\)
with intensities \( \mu^b \in \{0, \nu^b, \mu^b\} \).
In this study, we set \(D = 16\), with \(0 < \nu^a = \nu^b < \mu^a = \mu^b < 1\).
Here, \(\mu\) represents the signal state, \(\nu\) the decoy state, and
\(D\) the number of phase slices in phase randomization.

Bob then send their coherent states to a central measurement node (Charlie),
who performs interference measurements using a balanced beam splitter and two
single-photon detectors (left and right). Successful detection events are
announced, and rounds with a single detector click (either left or right) are
post-selected for pairing. Alice and Bob determine the basis and raw key values
for each pair based on relative intensity, phase information, and the frequency
difference acquired. Finally, they perform
parameter estimation, information reconciliation, and privacy amplification to
generate the final key. The length of the final key is given by~\cite{zeng2022mode}:
\begin{equation}
	K = M_{11}^{Z}\!\left[1 - h\!\left(e_{11}^{Z,\mathrm{ph}}\right)\right]
	- f \!\sum_{\mu} M_{\mu}^{Z}\,
	h\!\left(E_{\mu}^{Z}\right).
\end{equation}

In this formulation, \( h(x) \) denotes the binary entropy function,
\( f \) represents the error correction efficiency,
\( M^{Z}_{11} \) is the single-photon Z-pair count, and
\( e_{11}^{Z,\mathrm{ph}} \) signifies the phase error rate.
The key rate is \( R = K / N_{\mathrm{pair}} \), where \( N_{\mathrm{pair}} \)
represents half the total number of rounds.
A more detailed description of these steps can be found in the
Appendix.

\section{Experimental setup}
In our experimental setup, depicted in Fig.~\ref{fig:expsetup}, we employ a commercial narrow-linewidth laser operating at 1550 nm with an approximate linewidth of 1 kHz. The laser output is divided into two beams: one designated for frequency-matching and the other for quantum light interference. The quantum light beam first passes through an intensity modulator, converting the continuous wave into pulsed light at a repetition rate of 500 MHz and a pulse width of 375 ps. Subsequently, this pulsed light enters a Sagnac loop configuration—comprising a circulator, beam splitter, phase modulator, and optical fibers—which further modulates the light intensity to produce the desired decoy states\cite{post1967sagnac}. The light exiting the Sagnac loop is then phase-randomized by a phase modulator before being transmitted from the sender’s end.

After traversing the long optical fiber spool, the two quantum light beams reach Charlie, where polarization alignment is achieved using two polarization feedback modules. Each module comprises an electric polarization controller, a polarization beam splitter, and a single-photon detector. By adjusting the voltage of the electric polarization controller based on the photon counts from the single-photon detectors, one single-photon detector’s count is minimized, enhancing the interference contrast between Alice and Bob. The detectors monitoring the two paths post-interference at the beam splitter have a detection efficiency of 63\% and a dark count rate of 30 Hz. The optical loss at Charlie corresponds to an attenuation of 1.91 dB.

In the frequency-matching light interference process, the optical signals from Alice and Bob undergo the optical attenuation, serving as an alternative to optical fibers. This substitution is justified by the fact that the fiber in the frequency compensation channel has a negligible impact on frequency measurement. The rationale for using attenuation as a substitute for optical fiber is provided in the Appendix. This method enables rapid frequency difference measurements, typically achieving results within tens of nanoseconds. At Charlie’s side, the output light from the BS is detected by the photodiodes (Thorlabs PDB480C), converted into an electrical signal, and subsequently processed for high-precision timing processing.

We present the optimized intensity parameters for distances of 197.91 km and 296.80 km in Table.~\ref{tab:optimized_parameters}. For the X-basis, we measured the temporal delay between the signals from the single-photon detector and the photodiode for two fiber lengths, correlated the classical and quantum light interference results, and calculated the error rate. For the Z-basis, we achieved a high signal-to-noise extinction ratio using the intensity modulator and Sagnac ring, as evidenced by the observed error rate. A detailed description of the experimental procedure is provided in the Appendix.

\begin{table}[h]
	\centering
	\caption{\justifying Optimized intensity parameters. $\mu,\nu$ are the average photon numbers for signal and decoy; $p_\mu,p_\nu$ are the sending probabilities; $f$ is the error-correction factor~\cite{tang2023polar}.}
	\label{tab:optimized_parameters}
	\renewcommand{\arraystretch}{1.15}
	\resizebox{0.85\linewidth}{!}{
		\begin{tabular}{lcc}
			\toprule
			Parameter & 197.91 km & 296.80 km \\
			\midrule
			Channel loss (dB)   & 38.83 & 57.36 \\
			Sent pulses         & $2.08 \times 10^{12}$ & $9.86 \times 10^{12}$ \\
			$\mu$               & 0.3958 & 0.3591 \\
			$\nu$               & 0.0275 & 0.0232 \\
			$p_{\mu}$           & 0.15 & 0.25 \\
			$p_{\nu}$           & 0.25 & 0.30 \\
			$f$                 & 1.06 & 1.06 \\
			\bottomrule
		\end{tabular}
	}
\end{table}

\begin{table*}[htp!]
	\centering
	
		\caption{\justifying{Comparison of phase compensation approaches.}
		SNSPD: superconducting nanowire single-photon detector.
		Frequency reference: optical frequency reference.
		The table reports the X-basis error rate ($E_x$) for experiments at around $200$~km fiber.}
	\begin{tabular}{lcccc}
		\hline
		{Experiment} & {Method} & {SNSPD} & {Frequency reference} & {$E_x$} \\
		\hline
		Zhou et al.~\cite{zhou2023experimental} & High-finesse cavities & Not required  & Required & 26.94\% \\
		Zhang et al.~\cite{zhang2025experimental} & Data post-processing & Required & Not required  & 27.77\% \\
		Shao et al.~\cite{shao2025high} & Data post-processing & Required & Not required  & 28.91\% \\
		This work & Classical photodiode & Not required  & Not required  & 26.92\% \\
		\hline
	\end{tabular}

	\label{tab:phase}
\end{table*}

\section{Experimental results}

Under the 197.91~km fiber condition, we achieved a secure key rate of $6.11\times10^{-6}$, with a maximum pairing time of $10~\mu\text{s}$. The X-basis bit error rate was 26.92\% for the pairing time of $10~\mu\text{s}$, slightly exceeding the theoretical minimum of 25\% required for key generation. For the 296.80~km fiber, we extended the maximum pairing time to $20~\mu\text{s}$, yielding a key generation rate of $2.45\times10^{-7}$, which surpasses the PLOB bound of $1.20\times10^{-7}$. At this distance, the X-basis error rate was 27.76\% for the pairing time of $20~\mu\text{s}$.

Fig.~3 illustrates the temporal fluctuations in error rates for both the X- and Z-basis across two different fiber lengths. For the X-basis, the theoretical minimum error rate is 25\%. The observed system error arises from modulation-induced errors and phase fluctuation introduced by the fiber, which become more pronounced as the fiber length increases. A primary advantage of our frequency-matching method is that the measured phase error between the two matched pulses is predominantly determined by the phase difference arising from frequency mismatch, which occurs on a timescale of tens of nanoseconds. This implies that the frequency difference can be obtained quickly and accurately. Details are provided in the Appendix.


\begin{figure}[htbp]
	\includegraphics[width=8.5cm]{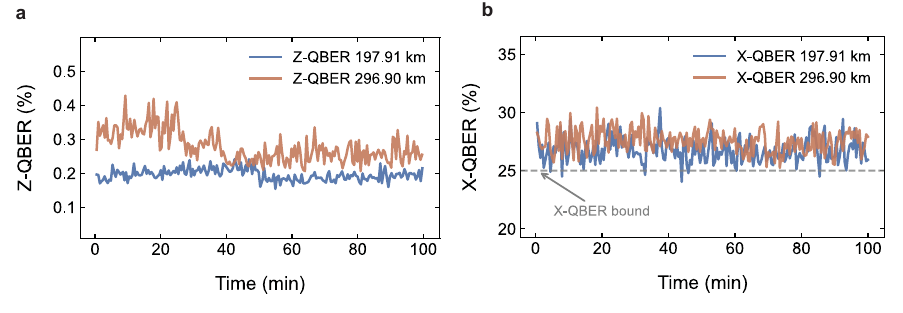}
	\caption{\justifying QBER drift. (a) Z-basis QBER maintains low values across the entire 100 min measurement, indicating stable quantum state preparation and detection. (b) X-basis QBER comparison under 197.91 km and 296.90 km. The X-QBER bound is indicated, demonstrating accurate phase compensation.}
	\label{fig:errdrift}
\end{figure}

Ideally, the theoretical error rate for the Z-basis can be reduced to zero. However, in practical implementations, errors primarily originate from incomplete extinction, dark counts, and other light noise. In our setup, the error rate on the Z-basis is consistently maintained below 0.3\%.

Additionally, we examine how the X-basis error rate varies with the maximum pairing length, as depicted in Fig.~\ref{fig:errdrift}. When the pairing length approaches 0, the dominant error contribution arises from device imperfections.
 As the pairing length increases, the error rate correspondingly rises, primarily due to phase fluctuations induced by fiber perturbations over the pairing duration.

\begin{figure}[htbp]
	\centering
	\includegraphics[width=8cm]{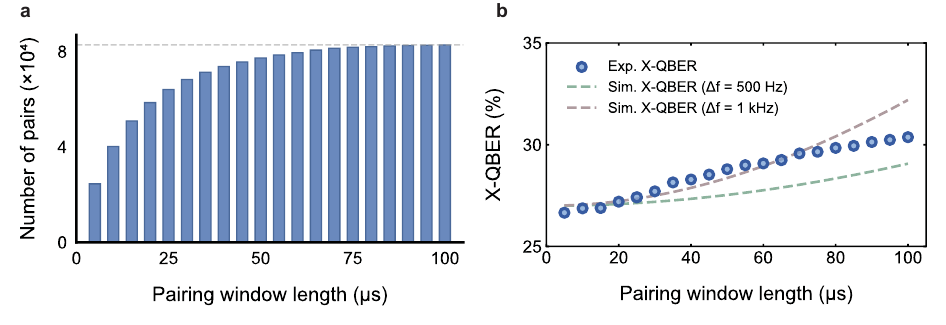}
	\caption{\justifying Total number of paired detection events and X-basis QBER as a function of the pairing window length under 296.80 km. (a) Total number of pairing events as a function of the pairing window length. A wider temporal window yields more matched detection events and gradually approaches saturation.
		(b) Experimental X-basis QBER (markers) versus pairing window length, together with simulated QBER under frequency detuning of   = 500 Hz and   = 1 kHz (dashed lines) confirming the precision of frequency compensation.}
	\label{fig:XerrwithLmax}
\end{figure}

We further examine the variation in the number of paired events under the mode-pairing QKD protocol at a transmission distance of 296.80~km, as Fig.~\ref{fig:XerrwithLmax} shown. Notably, when the maximum pairing length extends to $50~\mu\text{s}$, the number of pairings begins to saturate. Considering both the pairing rate and error performance, we compare our results with previous mode-pairing QKD experiments, as summarized in Table.~\ref{tab:phase}. Aside from this work, mode-pairing QKD implementations that exceed the rate-transmittance linear bound predominantly utilize optical cavities, commercial acetylene-stabilized laser systems, or data post-processing based on high-count-rate single-photon detectors. Our method offers a more practical and experimentally accessible implementation. Moreover, it achieves an X-basis error rate approaching the fundamental limit.

\section{Conclusion}

In summary, we develop a precise and robust scheme for stabilizing the frequency difference between two independent lasers and successfully implement it in a mode-pairing QKD system. our approach eliminates the need for high-fineness optical cavities, commercial acetylene-stabilized lasers, and single-photon detector-based data post-processing, while enabling seamless integration with classical communication channels.

 Moreover, incorporating an erbium-doped fiber amplifier into the frequency-matching optical path can extend communication distances by compensating for transmission losses~\cite{liu2021field}, enhancing its feasibility for field deployment. Additionally, employing multi-wavelength multiplexing in conjunction with this technique for mode-pairing QKD can further improve the key generation rate. By significantly reducing system complexity and cost, this method provides a practical and efficient framework for mode-pairing QKD, demonstrating strong potential for large-scale deployment in quantum networks. Furthermore, it offers a promising approach for integration into satellite-ground links, expanding its applicability in quantum communications.

\begin{acknowledgments}
This work is supported by the National Research Foundation,
Singapore and A*STAR under its Quantum Engineering Programme
(NRF2021-QEP2-01-P02, NRF2021-QEP2-04-P01, NRF2022-QEP2-02-P13),
A*STAR (M21K2c0116, M24M8b0004),
Singapore National Research Foundation (NRF-CRP22-2019-0004,
NRF2023-ITC004-001, NRFCRP30-2023-0003, NRF-MSG-2023-0002),
Singapore Ministry of Education Tier 2 Grant (MOE-T2EP50222-0018),
National Natural Science Foundation of China Grant No.~12174216.
We thank the support from Dieter Schwarz Stiftung gGmbH.
\end{acknowledgments}

\appendix

\section{Frequency matching method}

\begin{figure}[htbp]
	\centering
	\includegraphics[width=8cm]{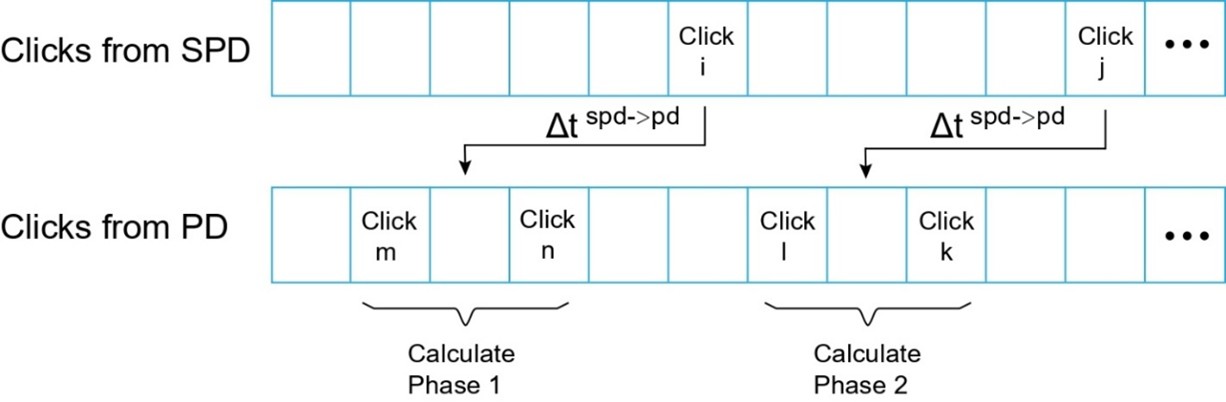}
	\caption{\justifying Schematic of phase calculation using the frequency-matching method. Detection events from the SPD are correlated with the corresponding clicks on the PD channel based on the detector response time difference. The clicks m, n, l, and k are used to compute the phase associated with clicks i and j, enabling continuous phase tracking}
	\label{fig:fmmethod}
\end{figure}

Here, we provide a detailed explanation of the methodology as Fig.~\ref{fig:fmmethod} shows.

To begin, pulsed light is employed to measure arrival times using three key components: 
a photodiode (PD), a single-photon detector (SPD), and a time-to-digital converter (TDC).
The TDC enables precise alignment of the time delay difference between PD and SPD detections.
By recording the arrival time of the pulsed light, we define the detection time at the PD as
$t_0^{\mathrm{PD}}$ and at the SPD as $t_0^{\mathrm{SPD}}$.
The temporal offset between these two detections, denoted as $\Delta t_0$, is given by:
\begin{align}
	\Delta t_0 = t_0^{\mathrm{SPD}} - t_0^{\mathrm{PD}}.
\end{align}

Next, we analyze the phase difference between two paired pulses in MP-QKD. 
Despite the fiber-induced effects, the phase difference can be expressed following the formulation~\cite{zhu2023experimental} as:
\begin{align}
	\Delta \theta_{i,j} = \int_{t_i}^{t_j} \omega(t)\, dt.
\end{align}

This expression involves an integral representation of the phase accumulation.
However, in practical scenarios, slight frequency fluctuations occur within the time interval between two detection events, introducing potential errors in phase estimation.
Consequently, rather than relying solely on integration, we compute the phase difference by considering the
arrival times of PD detection events that correspond to the response windows near the $i$-th and
$j$-th detections. These detection events are associated with specific points ($m,n,j,k$) in the
frequency matching process. Accordingly, the phase difference between the two paired pulses
is refined as follows:
\begin{align}
	\Delta\theta_{i,j} &= \theta_{j} - \theta_{i} \notag \\
	&= \frac{t_{j} - \Delta t_{0} - t_{i}}{t_{k} - t_{i}}
	\cdot \frac{t_{i} - \Delta t_{0} - t_{m}}{t_{n} - t_{m}}
	+ 2\pi N.
\end{align}

This formulation incorporates timing variations to provide a more accurate determination of
the phase difference between paired pulses, which is essential for precise phase matching in
MP-QKD. Here, $t_{i}$ and $t_{j}$ denote the TDC measurement times associated with the
MP-QKD paired detection events at positions $i$ and $j$. As illustrated in Fig.~\ref{fig:fmmethod}, the indices
$m,n,l,k$ correspond to the nearest PD detection events relative to clicks $i$ and $j$,
respectively, after compensating for the timing offsets between the SPD and PD detections.
The parameter $N$ represents the integer number of cycles that may occur between these
detection events. Importantly, since $N$ does not influence the final calculation of the
detected phase value, its precise value is not required.

\section{Fast Fourier transform (FFT) method}

As a comparison to the PD-based method, we also include in this work a method that
employs FFT-based phase estimation. The detailed comparison results are presented in the
 following text. In the FFT-based approach, the phase is estimated from the accumulated
detection responses of the superconducting detectors over a certain time window, based on the
detection probabilities of the detectors at the two output ports of the interference beam splitter:
\begin{align}
	P_{L}(t) &= \frac{1 + \cos\theta(t)}{2}, \\
	P_{R}(t) &= \frac{1 - \cos\theta(t)}{2}.
\end{align}

Here, $\theta(t) = 2\pi \Delta f\, t + \theta_{0}$.
Then we take the detection time as the horizontal axis, assign a value of $+1$ to detection
events at detector $L$ and $-1$ to those at detector $R$, and perform an FFT on the accumulated
data to obtain the frequency difference.

\section{Mode-pairing scheme}

\label{app:mpscheme}

Below, we present the steps of the mode-pairing quantum key distribution scheme with decoy-state method.

\textbf{1. State preparation:}
Alice and Bob split the continuous-wave (CW) light emitted from their respective lasers into two components.
One component remains unmodulated to facilitate frequency matching, following the methodology referenced in the following section.
The other component is individually modulated by Alice and Bob for mode-pairing QKD.
For the $i$-th pulse of quantum light, Alice prepares coherent state 
$\left|\sqrt{\mu_i^{A}}\, e^{\mathrm{i}\phi_i^A}\right\rangle$
on the optical mode $A_i$, with intensity $\mu_i^{A}$ chosen from $\{0,\nu,\mu\}$ 
($0<\nu<\mu<1$) randomly, and phase $\phi_i^A$ uniformly chosen from
$\left\{0,\frac{2\pi}{D},\frac{4\pi}{D},\ldots,\frac{2\pi(D-1)}{D}\right\}$,
with $D=16$.
Similarly, Bob randomly chooses $\mu_i^{B},\phi_i^{B}$ and prepares
$\left|\sqrt{\mu_i^{B}}\, e^{\mathrm{i}\phi_i^B}\right\rangle$ on mode $B_i$.
Here, $\mu$ represents the signal state, $\nu$ the decoy state, and $D$ the number of phase slices in phase randomization.

\textbf{2. State transmission and measurement:}
Alice and Bob send their pulses through optical modes $A_i$ and $B_i$, respectively, to an intermediate measurement node operated by Charlie.
Charlie performs single-photon interference measurements and announces detection events at either of the two detectors, labeled L and R.

\textbf{3. Repetition:}
This process (state preparation, transmission, and measurement) is repeated until each party transmits $N$ QKD pulses.
After transmission, Alice and Bob proceed with data post-processing.

\textbf{4. Mode pairing:}
For all rounds where a successful detection event (i.e., an L or R click) occurs, Alice and Bob apply the pairing strategy~\cite{zhu2023experimental} to associate two detected pulses as a pair.

\textbf{5. Basis assignment:}
Alice and Bob organize the paired QKD pulses into data pairs according to their intensities, following the categorization defined in Table~S1.
For a paired $i$-th and $j$-th round, the corresponding pulse intensities are denoted as $\mu_{i}^{A(B)}$ and $\mu_{j}^{A(B)}$, respectively.

\begin{table}[h!]
	\centering
	\caption{\justifying Basis assignment depending on pulse intensities.}
	\begin{tabular}{c|c|c|c}
		\hline
		$\mu_i$ $\backslash$ $\mu_j$ & 0 & $\nu$ & $\mu$ \\
		\hline
		0 & 0 & Z & Z \\
		\hline
		$\nu$ & Z & X & discard \\
		\hline
		$\mu$ & Z & discard & discard \\
		\hline
	\end{tabular}
	\label{tab:S1}
\end{table}

\textbf{6. Basis sifting:}
Alice and Bob announce the basis choices for all data pairs and the sum of their respective intensities, denoted as
$\tilde{\mu}_{B} = \left(\mu_{i}^{A} + \mu_{i}^{B},\, \mu_{j}^{A} + \mu_{j}^{B}\right)$
for each pair.
Subsequently, they further refine the basis selection of each data pair according to Table~\ref{tab:S2}.

\begin{table}[h!]
	\centering
	\caption{\justifying Basis sifting depending on intensity information from both sides.}
	\begin{tabular}{c|c|c|c|c}
		\hline
		Alice / Bob & 0 & Z & X & discard \\
		\hline
		0 & 0 & Z-pair & Z-pair & discard \\
		\hline
		Z & Z-pair & Z-pair & discard & discard \\
		\hline
		X & X-pair & discard & X-pair & discard \\
		\hline
		discard & discard & discard & discard & discard \\
		\hline
	\end{tabular}
	\label{tab:S2}
\end{table}

\textbf{7. Key mapping:}
For each Z-pair at location $(i,j)$, Alice assigns her key bit as:
\begin{equation}
	Z_A =
	\begin{cases}
		0, & \text{if } \mu_i^A = 0,\\
		1, & \text{if } \mu_i^A = \mu,
	\end{cases}
	\tag{S1}
\end{equation}

For X-pairs, the key bit is derived from the relative phase:
\begin{equation}
	X_A = \left( \frac{\phi_j^{A} - \phi_i^{A}}{\pi} \right) \bmod 2
	\tag{S2}
\end{equation}

Alice then announces the phase difference:
\begin{equation}
	\theta_A = (\phi_j^{A} - \phi_i^{A}) \bmod \pi
	\tag{S3}
\end{equation}

Similarly, Bob assigns his raw key bit $X_j^B$ and announces $\theta_B$.
The key assignment rules differ between Z-pairs and X-pairs:

{- For Z-pairs, Bob sets:}
\begin{equation}
	Z_B =
	\begin{cases}
		0, & \text{if } \mu_j^{B} = 0,\\
		1, & \text{if } \mu_j^{B} = \mu.
	\end{cases}
	\tag{S4}
\end{equation}

{- For X-pairs, Alice and Bob retain the key only if their relative phase difference satisfies:}
\begin{equation}
	\big|\theta_B - \theta_A\big| \in \left[ \Delta\theta_{i,j} + 2k\pi - \frac{\pi}{D}, \,
	\Delta\theta_{i,j} + 2k\pi + \frac{\pi}{D} \right]
	\tag{S5}
\end{equation}

where $k$ is an integer, and $\Delta\theta_{i,j}$ is determined from frequency matching according to Eq.~\ref{eq:phase_relation}.
If the condition is met, Alice and Bob keep the key; otherwise, they discard it.

\textbf{8. Parameter estimation:}
Alice and Bob utilize Z-pairs with different intensity settings to estimate the number of detected single-photon pairs 
$M_{11}^{Z}$ using the decoy-state method. Meanwhile, X-pairs are used to evaluate the single-photon phase error rate, following the post-processing procedure~[1].

\textbf{9. Key distillation:}
Alice and Bob utilize the $(\mu,\mu)$-pairs to generate secure keys.
The process consists of two key steps: error correction and privacy amplification, both based on the key rate formula incorporating parameters such as
$M_{11}^{Z}$, $e_{11}^{\mathrm{X}}$, $M_{\mu\mu}$ and $E_{\mu\mu}^{\mathrm{Z}}$.

The final key rate is determined as:
\begin{equation}
	M_R = M_{11}^{Z}\bigl(1 - h(e_{11}^{X})\bigr) - fM_{\mu\mu}E_{\mu\mu}
	\tag{S6}
\end{equation}

where $M_{11}^{Z}$ is the lower bound on the single-photon component in the Z-basis pairs and $e_{11}^{X}$ is the upper bound of the phase error rate in the X-basis.
These parameters are estimated using the decoy-state method and the Chernoff–Hoeffding bound, ensuring reliable security analysis even with finite-size effects.

\section{Discussion of frequency difference compensation and comparison with FFT method}

Here, we present an additional experiment under the frequency-matching method to investigate
how the fiber length affects the accuracy of frequency-difference estimation.
Furthermore, we provide another supplementary experiment that compares our method with the FFT-based
approach using post-processed SPD data.

In the X-basis, the phase difference between Alice and Bob for the $i$-th and $j$-th pulses in mode-pairing QKD
is given by~\cite{zhu2023experimental}:
\begin{equation}
	\Delta\theta_{i,j}
	= \Delta\theta_{0}^{A}
	+ (\Delta\theta_{i}^{B} - \Delta\theta_{i}^{A})
	+ \Delta\omega\cdot (t_j - t_i),
	\tag{S7}
	\label{eq:phase_relation}
\end{equation}

The optical fiber in the frequency-difference compensation channel only affects the third term—the
frequency difference. Considering a phase fluctuation rate of approximately $6.0~\mathrm{rad}/\mathrm{ms}$ 
over $550~\mathrm{km}$ fiber~\cite{lucamarini2018overcoming},
this corresponds to a beat frequency fluctuation of about $0.95~\mathrm{kHz}$.
Under a maximum pairing duration of $20~\mu\mathrm{s}$, this results in a maximum phase deviation of:
\begin{equation}
	0.95~\mathrm{kHz} \times 20~\mu\mathrm{s} \times 2\pi
	\approx 0.12~\mathrm{rad}
	< \frac{2\pi}{16},
	\tag{S8}
\end{equation}
which is negligible in terms of error rate.

\begin{figure}[htbp]
	\centering
	\includegraphics[width=8cm]{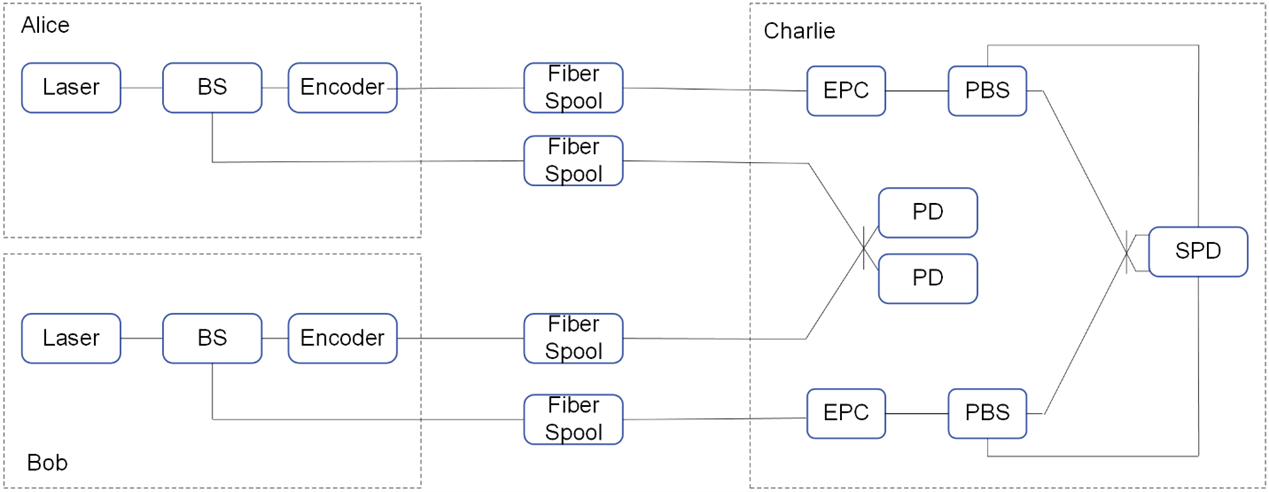}
	\caption{\justifying Experimental test using optical fiber as the frequency-matching channel.}
	\label{fig:compare}
\end{figure}

To further support this, we provide an experimental test of how the frequency estimation error changes
when $200~\mathrm{km}$ and $300~\mathrm{km}$ of fiber are added to the frequency-matching channel in
Table~\ref{tab:S3}. As shown in Fig.~\ref{fig:compare}, we additionally inserted $200~\mathrm{km}$ and $300~\mathrm{km}$ of optical
fiber into the frequency-matching channel, with corresponding $200~\mathrm{km}$ and $300~\mathrm{km}$ fiber
added to the QKD channel as well.

\begin{figure}[htbp]
	\centering
	\includegraphics[width=8cm]{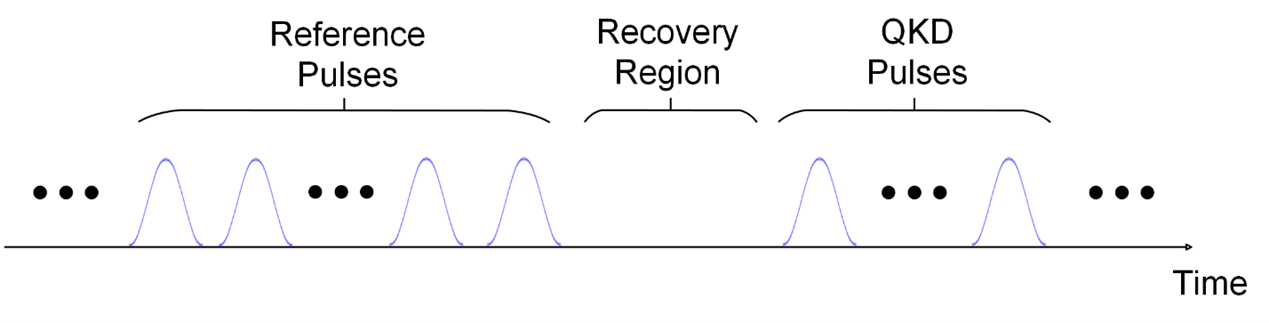}
	\caption{\justifying Emission timing of quantum light in the comparison experiment between frequency matching and FFT methods.}
	\label{fig:emission}
\end{figure}

Therefore, the fiber length in the frequency compensation channel has a negligible impact on the results.
This also confirms the validity of using attenuation as a simplification to replace optical fiber in
the frequency-matching channel in the main text.

In addition, we conducted another supplementary experimental test to compare the FFT-based method~\cite{zhang2025experimental,shao2025high}
with our frequency-matching approach under identical experimental conditions.

The setup remains the same as described in Fig.~\ref{fig:compare}, where a 300~km fiber spool was used.
In this test, the reference light was retained during emission at the sender’s side, such that the quantum
signals from Alice and Bob were mixed with the reference light. The emission timing is shown in Fig.~\ref{fig:emission}.

\begin{table}[h!]
	\centering
	\caption{\justifying X-basis error after adding fiber to the frequency matching channel.}
	\begin{tabular}{c|c|c}
		\hline
		& 200 km & 300 km \\
		\hline
		Max pairing length & 20~$\mu$s & 20~$\mu$s \\
		\hline
		Total pairs       & 102745 & 96717 \\
		\hline
		Error pairs       & 27665 & 26053 \\
		\hline
		Error rate                     & 26.92\%  & 26.93\% \\
		\hline
	\end{tabular}
	\label{tab:S3}
\end{table}

Here, among the 500~MHz emission frequency of the quantum light, 300~MHz is allocated to
reference light for FFT-based frequency difference estimation, with a reference light count rate
of 0.8~M clicks/s at Charlie’s side; 50~MHz is reserved as a recovery interval with no emission,
and 150~MHz is used for QKD. Meanwhile, 300~km of optical fiber is also added to the frequency-matching
channel for frequency matching.

As shown in Table~\ref{tab:S4}, our method achieves higher accuracy compared to the FFT-based approach~\cite{zhang2025experimental,shao2025high},
which shows that FFT requires a relatively high-count rate to maintain a sufficiently low error rate.

\begin{table}[h!]
	\centering
	\caption{\justifying Comparison between frequency matching and FFT method.}
	\begin{tabular}{c|c|c}
		\hline
		& Frequency Matching & FFT \\
		\hline
		Max pairing length & 20~$\mu$s & 20~$\mu$s \\
		\hline
		Total pairs       & 51619 & 50851 \\
		\hline
		Error pairs       & 13743 & 14049 \\
		\hline
		Error rate                     & 26.62\% & 27.62\% \\
		\hline
	\end{tabular}
	\label{tab:S4}
\end{table}

\section{Experimental details}

At the transmitter side, we employ a commercial narrow-linewidth laser (linewidth about
1~kHz) operating at 1550~nm to generate a continuous-wave optical source.
The emitted light is divided into two beams via a beam splitter (BS): one beam is allocated
for frequency matching, while the other undergoes intensity modulation combined with a
Sagnac loop to implement three decoy states.
Additionally, a phase modulator introduces random phase modulation, applying 16 distinct
phase values to the pulsed light.

The modulation signals are generated using an arbitrary waveform generator (Tektronix
AWG70002A), operating at 500~MHz.
The pulse widths of the electrical signals driving the modulators are 375~ps (intensity
modulation), 625~ps (Sagnac loop modulation), and 1~ns (random phase modulation).
To ensure synchronization, identical AWGs are employed for both Alice and Bob.

For the frequency-matching channel, an optical attenuator is used instead of a fiber spool to
achieve frequency matching with the photodiode detector (Thorlabs PDB480C-AC).
Given that the frequency offset for matching is maintained within 50~MHz, the presence or
absence of additional optical fiber has negligible impact on the frequency-matching outcome.
The quantum channel is implemented using standard optical fiber, with its total length and
attenuation detailed in the main text.

At the receiver side, an electric polarization controller (EPC) in conjunction with a polarizing
beam splitter (PBS) is utilized for polarization feedback control.
By appropriately adjusting the EPC voltage, the photon count in the non-interfering port of the PBS
is minimized, ensuring the polarization states of Alice and Bob remain aligned.
The two pulsed beams intended for interference undergo detection using superconducting nanowire
single-photon detectors (SNSPDs), and the resulting signals are processed by time-to-digital
converter (TDC) channels.

\begin{table}[htp!]
	\centering
	\caption{Detailed experimental results.}
	\begin{tabular}{l|c|c}
		\hline
		\hline
		& 197.97 km & 296.80 km \\
		\hline
		$\mu^A$ & 0.3985 & 0.3591 \\
		$\nu^A$ & 0.0275 & 0.0232 \\
		$p_d^A$ & 0.15 & 0.25 \\
		$p_s^A$ & 0.25 & 0.30 \\
		$\alpha$ & 0.1962 & 0.1936 \\
		$\eta^A$ & $2.87\times10^{-2}$ & $3.63\times10^{-4}$ \\
		$N^A$ & $2.08\times10^{12}$ & $9.86\times10^{12}$ \\
		$e_0^A$ & 0.1692 & 0.2363 \\
		$M_{11}^A$ & $2.181\times10^{7}$ & $8.222\times10^{6}$ \\
		QBER & $1.979\times10^{-3}$ & $2.520\times10^{-3}$ \\
		$T_{\max}$ & 10 $\mu$s & 20 $\mu$s \\
		\hline
		Sent-$A_{0}B_{0}$   & 749131506000  & 1998523398375 \\
		Sent-$A_{0}B_{\mu}$ & 3121381275000 & 1332348932250 \\
		Sent-$A_{\mu}B_{0}$ & 1872828765000 & 1110290776875 \\
		Sent-$A_{\mu}B_{\mu}$ & 3121381275000 & 1332348932250 \\
		Sent-$A_{0}B_{0}$   & 130057553125  & 888232621500 \\
		Sent-$A_{0}B_{\mu}$ & 780345381875  & 740193851250 \\
		Sent-$A_{\mu}B_{0}$ & 1872828765000 & 1110290776875 \\
		Sent-$A_{\mu}B_{\mu}$ & 780345381875 & 740193851250 \\
		\hline
		Detected-$A_{0}B_{0}$   & 847918       & 346662 \\
		Detected-$A_{0}B_{\mu}$ & 24696156     & 11143558 \\
		Detected-$A_{\mu}B_{0}$ & 212343799    & 145200335 \\
		Detected-$A_{\mu}B_{\mu}$ & 24736097   & 11394242 \\
		Detected-$A_{0}B_{0}$   & 20400927     & 14858608 \\
		Detected-$A_{0}B_{\mu}$ & 94640175     & 102896817 \\
		Detected-$A_{\mu}B_{0}$ & 213546450    & 142398321 \\
		Detected-$A_{\mu}B_{\mu}$ & 95062049   & 102559318 \\
		Detected-$A_{0}B_{\mu}$ & 106401117    & 160983600 \\
		\hline
		\hline
	\end{tabular}
	\label{tab:S5a}
\end{table}

To enhance interference visibility, a 360~ps time window is set, ensuring precise temporal
separation between interference events.
Concurrently, the frequency matching detection signal is linked to a separate TDC channel.
This approach enables accurate determination of the frequency difference near any individual
detection response, facilitating X-basis error rate estimation and overall parameter optimization.
The experimental setup yields verified results for fiber transmission distances of 197.91~km
and 296.80~km, as presented in Table~\ref{tab:S5a} and Table~\ref{tab:pairdetails}.

\begin{table}[htp!]
	\centering
	\caption{Pair details.}
	\begin{tabular}{c|c|c}
		
		\hline
		\hline
		Distance                        & 197.97 km           &  296.80 km            \\ \hline
		$ \textit{Z}_{A0B0} $-\textit{Total}       & 417        & 66        \\ 
		$ \textit{Z}_{A0B0} $-\textit{Error}       & 417        & 66       \\  
		$ \textit{Z}_{A0B\nu} $-\textit{Total}        & 24195        & 3510       \\  
		$ \textit{Z}_{A0B\nu} $-\textit{Error}        & 11954        & 1758       \\  
		$ \textit{Z}_{A0B\mu} $-\textit{Total}      & 207553        & 44887       \\  
		$ \textit{Z}_{A0B\mu} $-\textit{Error}      & 103944         & 22294       \\  
		$ \textit{Z}_{A\nu B0} $-\textit{Total}        & 24313        & 3504       \\  
		$ \textit{Z}_{A\nu B0} $-\textit{Error}        & 12211       & 1715       \\  
		$ \textit{Z}_{A\nu B\nu} $-\textit{Total}        & 720585        & 118843       \\  
		$ \textit{Z}_{A\nu B\nu} $-\textit{Error}      & 19905        & 4611       \\  
		$ \textit{Z}_{A\nu B\mu} $-\textit{Total}      & 6120259         & 1528906       \\  
		$ \textit{Z}_{A\nu B\mu} $-\textit{Error}        & 92792        & 32028        \\  
		$ \textit{Z}_{A\mu B0} $-\textit{Total}        & 208955        & 44650       \\  
		$ \textit{Z}_{A\mu B0} $-\textit{Error}        & 104453       & 22302       \\  
		$ \textit{Z}_{A\mu B\nu} $-\textit{Total}        & 6137758        & 1521102       \\  
		$ \textit{Z}_{A\mu B\nu} $-\textit{Error}      & 92658        & 31359       \\  
		$ \textit{Z}_{A\mu B\mu} $-\textit{Total}      & 52138562         & 19499824       \\  
		$ \textit{Z}_{A\mu B\mu} $-\textit{Error}        & 103219        & 49148        \\  \hline
		
		$ \textit{X}_{A0 B2\nu} $-\textit{Total}      & 353629         & 58026       \\  
		$ \textit{X}_{A0 B2\nu} $-\textit{Error}        & 177809        & 28919        \\  
		$ \textit{X}_{A0 B2\mu} $-\textit{Total}      & 26111151         & 9851306       \\  
		$ \textit{X}_{A0 B2\mu} $-\textit{Error}        & 13126550        & 4895936        \\       
		$ \textit{X}_{A2\nu B0} $-\textit{Total}      & 355455         & 56967       \\  
		$ \textit{X}_{A2\nu B0} $-\textit{Error}        & 179256        & 28338        \\      
		$ \textit{X}_{A2\nu B2\nu} $-\textit{Total}      & 29819         & 12500       \\  
		$ \textit{X}_{A2\nu B2\nu} $-\textit{Error}        & 8028        & 3470        \\              
		$ \textit{X}_{A2\nu B2\mu} $-\textit{Total}      & 645893         & 616871       \\  
		$ \textit{X}_{A2\nu B2\mu} $-\textit{Error}        & 286838        & 276913        \\          
		$ \textit{X}_{A2\mu B0} $-\textit{Total}      & 26488584         & 9666134       \\  
		$ \textit{X}_{A2\mu B0} $-\textit{Error}        & 13348620        & 4803449        \\      
		$ \textit{X}_{A2\mu B2\nu} $-\textit{Total}      & 654816         & 606464       \\  
		$ \textit{X}_{A2\mu B2\nu} $-\textit{Error}        & 290939        & 272583        \\              
		$ \textit{X}_{A2\mu B2\mu} $-\textit{Total}      & 816963         & 1498977       \\  
		$ \textit{X}_{A2\mu B2\mu} $-\textit{Error}        & 219169        & 411379        \\    
		\hline
		\hline
	\end{tabular}
	
	    \label{tab:pairdetails}
	\end{table}

\newpage


%



\end{document}